\input epsf
\documentstyle[12pt,aasms4]{article}

\received{23 October 1995}
\accepted{9 April 1996}

\begin{document}

\title{On the Possible Variations of the Hubble Constant with Distance}

\author{Xiang-Ping Wu}
\affil{Department of Physics, University of Arizona, Tucson, AZ 85721 and \\
       Beijing Astronomical Observatory, Chinese Academy of Sciences,
       Beijing 100080, China}

\author{Bo Qin}
\affil{Beijing Astronomical Observatory, Chinese Academy of Sciences,
       Beijing 100080, China}

\and

\author{Li-Zhi Fang}
\affil{Department of Physics, University of Arizona, AZ 85721}

\begin{abstract}
Current measurements of the Hubble constant $H_0$ on scale 
less than $\sim100$ Mpc appear to be controversial, while 
the observations made at high redshift seem to provide a 
relatively low value. 
On the other hand, the  Hubble expansion is
driven by the matter content of the universe. 
The dynamical analysis on scale of 
a few $\sim10$ Mpc indicates that the  matter density
$\Omega_0$ is only $\sim0.2$--$0.3$, which is 
significantly smaller than $\Omega_0=1$ predicted
in the standard inflation model. This might support
the tendency of a decreasing Hubble constant towards distance.
In this paper, we 
discuss the influence of a possible variant Hubble constant on
two fundamental relations in astronomy: the 
magnitude-redshift ($m$--$z$)
and the number-magnitude relations.
Using a distant type Ia supernova at $z=0.458$,
we show that the deceleration parameter $q_0$ or 
$\Omega_0$ cannot be determined from the 
$m$--$z$ relation at moderate/high redshift unless the
variation of the Hubble constant is {\it a priori} measured. 
It is further demonstrated that the number density
of distant sources would be underestimated when 
their local calibration is employed, which may partially
account for the number excess of the faint blue galaxies
observed at moderate/high redshift.
\end{abstract}

\bigskip


\section{Introduction}

The Hubble constant $H_0$ is an indicator of the global expansion 
of the universe.  Theoretically, this parameter is defined in terms
of the Cosmological Principle, i.e.,  the overall matter 
distribution of the universe is assumed to be isotropic and 
homogeneous.
The isotropy of the universe has been remarkably demonstrated 
by the measurement of the 
$3^{\circ}$K microwave background radiation 
(e.g. Bennett et al. 1996). 
However, observations have indicated that there exists an 
inhomogeneous matter distribution
in the local universe on scale up to $\sim100$ Mpc, 
such as matter clumps, voids, filaments, etc., 
which may cause a deviation of the local Hubble expansion  
from the global one. The pioneering work to directly measure the
variation of the Hubble constant with distance
was carried out  by Sandage, Tammann and Hardy in 1972, 
using the first ranked E galaxies.  Two decades 
later, Lauer \& Postman (1992) conducted another direct 
observation from the brightest cluster galaxies to $z=0.05$.
These authors have essentially reached a similar conclusion 
that the ratio of the local Hubble constant $H_L$ to the
global one $H_G$ is consistent with unity, 
indicative of a very minor effect of 
local matter clumps and voids on the expansion of the universe.
Other evidences from the study of the Hubble diagram of various
objects (e.g. Sandage \& Hardy 1973; Tammann \& Sandage 1995)   
supported that the Hubble constant is roughly invariant.

However, this situation has been recently challenged by several  
determinations of the Hubble constant utilizing 
Cepheid variables and the globular cluster luminosity 
function in some nearby galaxies, which yield  
a  large value of  $H_0$ ranging from
69 to 87 km s$^{-1}$ Mpc
(Pierce et al. 1994; Freedman et al. 1994; 
Tanvir et al. 1995; Whitmore et al. 1995).
These correspond to an age of the universe 
of $8-13$ Gyr in the frame of the standard cosmological model,
while the oldest known stars in globular 
clusters of our Galaxy are estimated to be about 
$15.8\pm2.1$ Gyr (Bolte \& Hogan 1995).
If these calibrations are correct rather than suffer from
the systematic and logical errors as were argued by Sandage 
and collaborators
(Sandage 1996; Sandage \& Tammann 1996; reference therein),
there are two possibilities to remove the 
apparent conflict over the age of the universe: 
(1) a nonzero cosmological constant 
$\lambda_0\equiv\Lambda/(3H_0^2)$
[see Ostriker \& Steinhardt (1995) for a summary]
 and (2) a local low-density region 
embedded in a globally flat universe 
(Turner, Cen \& Ostriker 1992; Wu et al. 1995). 
The later requires a considerably variation of the 
Hubble constant with distance. 

The claim for a deviation of the local Hubble flow from
the global one is indirectly supported by the  following
observations: From the time delay of 1.1 years revealed 
by over 10 years coverage of the monitoring of the 
gravitationally-doubled images QSO0957+561A,B at redshift of 1.41, 
one has found $H_G=48^{+16}_{-7}$ km s$^{-1}$ Mpc$^{-1}$
[see Wu (1996) for a recent review]. 
This global value is relatively low though large uncertainties
may exist in the resulting $H_G$ due to the modeling of 
the deflectors. For instance, a new lensing model of 
QSO0957+561A,B based on the same data yields $H_G\approx
82.5$ km s$^{-1}$ Mpc$^{-1}$ (Grogin \& Narayan 1996).
Another evidence comes 
from the measurement of Sunyaev-Zel'dovich (S-Z) effect,
which is the spectral distortion of the cosmic 
background radiation due to the inverse Compton cooling of 
the hot X-ray gas in clusters of galaxies  
(Sunyaev \& Zel'dovich 1972).  
It has been found that a relatively small value of the Hubble
constant is derived from distant galaxy cluster, e.g. 
$H_0=65\pm25$ km s$^{-1}$ Mpc$^{-1}$ for A2218 ($z=0.171$) 
(Jones et al. 1993; Birkinshaw \& Hughes 1994),
$H_0=47\pm17$ km s$^{-1}$ Mpc$^{-1}$ for A665 ($z=0.182$)  and
$H_0=41^{+15}_{-12}$ km s$^{-1}$ Mpc$^{-1}$ for Cl0016+16 ($z=0.545$) 
(Yamashita 1994). However, the detection of the S-Z effect
for nearby Coma cluster provides 
$H_0=71_{-25}^{+30}$ km s$^{-1}$ Mpc$^{-1}$ (Herbig et al. 1995). 
It seems likely that the Hubble constant decreases with redshift.
Yet, the errors in these measurements are still large and 
uncertainties from modeling of the hot X-ray gas in clusters
of galaxies need to be further studied (Inagaki, Suginohara
\& Suto 1995; Rephaeli 1995). 

On the other hand, it is crucial to realize that 
the Hubble constant depends on the
matter distribution since the expansion is uniquely 
governed by the gravitational matter of the universe. 
There have been increasing observational evidences that 
the local matter density $\Omega_L$ within a few 10 Mpc is only 
$\Omega_L\approx0.2$--$0.3$ (Bahcall, Lubin \& Dorman 1995;
references therein). However, the standard inflation
cosmological model predicts an overall matter density of
$\Omega_G\approx1$. 
This matter discrepancy implies
a gradient of the Hubble constant along distance.

The present measurement of the Hubble constant 
as well as the density parameter 
is still a controversial issue. It turns out to be unclear
whether or not the Hubble constant alters with distance.
In this paper, we address
the following question: If the Hubble constant is variant,
what effect will it have on the fundamental relations 
in astronomy ?  In particular we will demonstrate this
effect on the magnitude-redshift relation using 
the distant type Ia supernova (section 2) and on the source 
number counts using the faint galaxy population (section 3).
In section 4 we will construct a simple analytic model for a 
locally low and globally high density universe 
instead of the usual simulation method (e.g. Turner et al. 1992;
Suto et al. 1995; Nakamura \& Suto 1995; Shi et al. 1995).
A brief discussion and our conclusions will be given in section 5.

\section{Comparison of nearby and distant supernovae}

Study of the magnitude-redshift relation ($m$--$z$), i.e. the Hubble
diagram, at moderate and high redshift has been thought to be
able to  provide
information on the deceleration parameter $q_0$ or $\Omega_0$. 
However, to do this one actually needs
to presume that $H_0$ is invariant with redshift. In the past 
decades many attempts have been made to derive $\Omega_0$ from 
the $m$--$z$ relation for various objects.
We will illustrate this procedure and discuss the influence of
the variation of $\Omega_0$ or $H_0$ on 
$m$--$z$ relation using the 
most distant type Ia supernova (SN Ia) discovered recently 
at $z=0.458$ (Perlmutter et al. 1995).

Because of the narrow distribution of their absolute magnitudes at 
maximum light, SN Ia's are often considered to be 
``standard candles'' and therefore, used as distance indicators 
(Branch \& Tammann 1992; Branch, Nugent \& Fisher 1996 and 
references therein). 
The detection of $\sim20$ distant SN Ia's at redshift 
$z\approx 0.35-0.5$ provides an unprecedented method of measuring 
the geometry of distant universe 
(Perlmutter et al. 1995; Goobar \& Perlmutter 1995; 
Branch et al. 1996). 
Neglecting the $K$ correction and the extinction of our 
Galaxy for the moment, the $m$--$z$ relation of
a SN Ia with absolute magnitude $M$ is
\begin{equation}
m=M+5\log (D_L/10{\rm pc}),
\end{equation}
where $D_L$ is the luminosity  distance:
\begin{equation}
D_L=\left(\frac{2c}{H_0}\right)
\frac{z\Omega_0+(\Omega_0-2)(-1+\sqrt{\Omega_0 z+1})}{\Omega_0^2}.
\end{equation}
If the absolute magnitude $M_{\rm max}$ at maximum light is 
calibrated using the nearby 
SN Ia's, e.g. for $B$ band (Perlmutter et al. 1995), 
\begin{equation}
M_{\rm max}-5\log(H_0/75)=-18.86\pm0.06,
\end{equation}
then the monitoring of the apparent magnitude variation 
of distant SN Ia's  
can give rise to the mean mass density of the universe, $\Omega_0$.
Calibrating the $B$ light curve of a distant SN Ia at $z=0.458$
by eq.(1) and the template standard model for nearby SN  Ia's,
Perlmutter et al. (1995) reached a result of $\Omega_0\approx0.2$.
At the time of the refereeing process, 
another $20$ distant SN Ia's 
have been found and a similar inverstigation has been made 
by the LBL group (Perlmutter et al. 1996; Kim et al. 1996), 
which leads to a new 
determination of $\Omega_0\approx1$.  
Note that their analysis was based on the hypothesis that the local
expansion rate $H_L$ remains as the same as 
the global Hubble flow $H_G$
so that the Hubble constant does not appear in their final result.

What would happen if $H_L\neq H_G$ ?
In a locally low but globally high density universe 
($H_L>H_G$), the observed apparent magnitude of a distant 
SN Ia at  maximum light should be generally written as
\begin{equation}
m_{\rm max}=\left(M_{\rm max}-5\log\frac{H_L}{75}\right)-
            5\log\frac{H_G}{H_L}
+5\log\frac{\Omega_0 z+(\Omega_0-2)(\sqrt{1+\Omega_0z}-1)}
{\Omega_0^2}+5\log\frac{2c}{75},
\end{equation}
where the first term on the right hand side is the local 
calibration at maximum light and the second one denotes   
the deviation of the local expansion from the global Hubble flow.
Compared with the conventional Mattig's relation,
eq.(4) contains two free parameters: $\Omega_0$ and
$H_L/H_G$. 
Taking the observational data 
from the first discovered distant SN Ia,
$m_{R,{\rm max}}=22.2$, $A_R=0.006$ ($R$ extinction) and 
$\bigtriangleup m_{RB}=-0.7$ (difference of $R$ and $B$ magnitudes) 
(Perlmutter et al. 1995), we have plotted the variation of $H_L/H_G$ 
versus $\Omega_0$ in Fig.1, together with 
the total errors introduced by the intrinsic 
dispersion of $M$ for the SN Ia ($\pm 0.25$) and other errors 
($\pm 0.16$) from the photo noise, calibration, etc. 
The major error stems from the intrinsic dispersion, 
which can be greatly
reduced with the employment of more distant SN Ia's.
Based on one distant  SN Ia at $z=0.458$, limit on 
the variation of the Hubble constant is 
$H_L/H_G=1.02^{+0.22}_{-0.17}$ and 
$1.12^{+0.23}_{-0.19}$ for
a global density of $\Omega_G=0.2$ and  $\Omega_G=1$, respectively. 
Because of the large errorbars, the variation of $H_L/H_G$ 
has not been well constrained. However, 
this example shows that the $m$--$z$ relation cannot be
used for the determination of the global $\Omega_{G}$ 
unless the variation of the Hubble constant is {\it a priori} given.

\placefigure{fig1}

\section{Comparison of nearby and distant galaxies}

The deviation of $H_L$ from $H_G$ also affects the 
calibration of the number-magnitude relation.
Compared with the local sources (e.g. galaxies), 
the luminosity distance $D_L$ 
and the volume element in the distant universe 
change by a factor of $H_L/H_G$ and 
$\sim (H_L/H_G)^3$, respectively.
As a result, the number density
of the sources at high redshift would appear to be
different from the local one.

We now show this effect by considering 
the galaxy number counts. We employ the Schechter 
luminosity function for the galaxy luminosity distribution
which has been calibrated locally (Efstathiou, Ellis 
\& Peterson 1988): $\phi^*=1.56\times10^{-2}\;h_L^3$
Mpc$^{-3}$, $\alpha=-1.11$ and $M^*=-19.6+5\log h_L$
where $h_L=H_L/100$ km s$^{-1}$ Mpc$^{-1}$. 
We keep the same galaxy morphological composition
as those used by Yoshii \& Sato (1992):
(E/S0,Sab,Sbc,Scd,Sdm)=(0.215,0.185,0.160,0.275,0.165)
and take the $K$ correction  from King \& Ellis (1985).
Because our goal is to demonstrate the influence of $H_L/H_G$,
we will not include the evolution correction which
is dependent of the cosmological models ($H_0$ and $q_0$).
Furthermore, we assume a constant 
comoving number density of galaxies and
extrapolate the above conditions up to $z=3$. 
Fig.2 shows the differential number counts of galaxies 
per square degree in $B$ band for
$H_L=H_G$ and $H_L=1.3H_G$, respectively,
where the global matter density is chosen to be unity. 
It turns out that there are no apparent differences 
in the curves at the bright limiting magnitude. 
However, a significant excess of faint population 
is predicted for $H_L>H_G$.
Note that the scale in the $m-$axis should not be regarded
as an accurate indicator of the magnitude due to
the ignorance of the color and evolution corrections.

\placefigure{fig2}

This number disprepancy of faint galaxies arising from the
local calibration of the Hubble constant may partially
explain the excess of number counts of faint blue galaxies
in observations (Broadhurst, Ellis \& Glazebrook 1992; 
references therein). Meanwhile, 
this might resolve the too few giant luminous arcs predicted
from the statistical lensing by clusters of galaxies, in which
background galaxies were assumed to
follow their local properties (Wu \& Hammer 1993; 
Wu \& Mao 1996).  Arcs/arclets have been shown to
be a very efficient tool for probe of the distant galaxies
[see Fort \& Mellier (1994) for a recent review]. 
Alternatively, the magnification probability of distant 
quasars by foreground galaxies (Turner, Ostriker \& Gott 1984) 
is increased by a factor of roughly $(H_L/H_G)^3$ while
the expected separation of images remains unchanged.

\section{Modeling of the local universe}

While the dynamical measurement shows $\Omega_L\approx0.2$--$0.3$
(Bahcall et al. 1995), we are situated in a low-density region
if $\Omega_G$ is close to unity as predicted by the standard
inflation model. We now construct a simple model for 
the local universe. 
The Tolman-Bondi metric is suitable for the description of the 
space-time of a spherical perturbation in an expanding universe.  
For an initial underdense perturbation $\delta_i$ at the cosmic 
epoch $t=t_i$ or at redshift $z_i$, its evolution follows
\begin{equation}
\frac{r}{r_i}=\frac{\Omega_i(1-\delta_i)}{1-\Omega_i(1-\delta_i)}
              \frac{\cosh\eta-1}{2};
\end{equation}
\begin{equation}
\sinh\eta-\eta=\frac{2[1-\Omega_i(1-\delta_i)]^{3/2}}{\Omega_i(1-\delta_i)}
                 \left(H_it_i\right)\left(\frac{t}{t_i}-1\right)+
               \frac{2\sqrt{1-\Omega_i(1-\delta_i)}}{\Omega_i(1-\delta_i)}-
            \cosh^{-1}\frac{2-\Omega_i(1-\delta_i)}{\Omega_i(1-\delta_i)},
\end{equation}
where $\Omega_i$ and $H_i$ are the mean mass density and the Hubble 
constant at $t=t_i$ for the background universe, respectively,
which are related 
to their present values through
\begin{equation}
\Omega_i=\frac{\Omega_0(1+z_i)}{1+\Omega_0z_i},
\end{equation}
\begin{equation}
H_i=H_0(1+z_i)\sqrt{1+\Omega_0 z_i}.
\end{equation}
For the sake of simplicity as well as the desire of the inflation model,
the background universe is often taken to be flat in the previous similar 
studies (e.g. Nakamura \& Suto 1995;
Wu et al. 1995). With the choice of $\Omega_0=1$ eqs.(5) and (6)
reduce to the results of Wu et al. (1995). 
We are now dealing with a global open/flat universe. 
The ages of an open universe at $t=t_i$ and $t=t_0$ (present) are 
\begin{equation}
t_i=\frac{1}{H_0}\left(\frac{\sqrt{1+\Omega_0z_i}}{(1-\Omega_0)(1+z_i)}-
    \frac{\Omega_0}{2(1-\Omega_0)^{3/2}}
    \cosh^{-1}\frac{2-\Omega_0+\Omega_0z_i}{\Omega_0(1+z_i)}\right)
\end{equation}
and
\begin{equation}
t_0=\frac{1}{H_0}\left(\frac{1}{1-\Omega_0}-
     \frac{\Omega_0}{2(1-\Omega_0)^{3/2}}
     \cosh^{-1}\frac{2-\Omega_0}{\Omega_0}\right),
\end{equation}
respectively. 
Defining the present Hubble constant as $H_0\equiv \dot{r_0}/{r_0}$, we
can find the relationship between the local Hubble constant $H_L$ and 
the background one $H_G$ to be
\begin{equation}
\frac{H_L}{H_G}=\frac{(1+z_i)\sqrt{1+\Omega_i z_i}}{(r_0/r_i)}
      \sqrt{1-\Omega_i(1-\delta_i)+\frac{\Omega_i(1-\delta_i)}{(r_0/r_i)}}.
\end{equation}
Finally, the energy conservation gives rise to the 
present mass density of the local universe $\Omega_{L}$ 
developed from the initial fluctuation $\delta_i$:
\begin{equation}
\Omega_{L}=\Omega_0(1-\delta_i)\left[\frac{1+z_i}{(r_0/r_i)}\right]^3.
\end{equation}

Setting the epoch of the initial perturbation of our local 
low-density region
to be the decoupling of $z_i\approx1000$, we 
have performed the numerical computations of $H_L/H_G$ 
versus $\Omega_0$ for the local mass density of 
$\Omega_{L}=0.1$, $0.2$ and $0.3$ and 
illustrated the results in Fig.3.  
As is expected, the deviation of 
the local expansion rate from the global Hubble flow 
depends critically on the mass density difference between 
the local and background universe. 
Using the local dynamical measurement of 
$\Omega_{L}\approx0.2$--$0.3$ 
(Bahcall et al. 1995), we can set the variation 
of the Hubble constant in a global flat universe to be
$H_L/H_G\approx1.3$.  Recall that $H_L/H_G=1.5$ in
the extreme case of where $\Omega_L=0$ and $\Omega_G=1$.

\placefigure{fig3}

\section{Discussion and conclusions}

We have shown that if the matter density of our local universe
within $\sim100$ Mpc is $\Omega_L\approx0.2$--$0.3$, as indicated
by the dynamical observations (Bahcall et al. 1995), and the global
value is close to unity, as predicted by the standard inflation
model, then the local Hubble expansion rate would be $\sim1.3$ times
larger than the global one. The current measurements of the 
Hubble constant do not exclude this variation with distance. 
As a consequence, we need to be cautious of using
some fundamental relations.

Utilizinging a distant SN Ia at $z=0.458$ discovered recently
(Perlmutter et al. 1995), we have demonstrated how the variation
of $H_L/H_G$ affects the magnitude-redshift 
relation.  Given the global matter density $\Omega_G$,
we have placed the preliminary limit on $H_L/H_G$. 
It turns out that due to
the large intrinsic dispersion for the SN Ia, the variation of 
$H_L/H_G$ has not been well constrained. The
current limit is consistent with either a uniform expansion
or a variation of $H_L/H_G$. Nevertheless, 
we have pointed out that there are two free parameters, 
$\Omega_G$ and $H_L/H_G$, appearing in the $m$--$z$ relation at
moderate/high redshift if the local calibration is used, 
which would invalidate the conventional
method of determining $\Omega_0$ from the difference in the Mattig
relation unless observations are able to provide 
the relationship between $H_L$ and $H_G$. This is consistent
with the recent work by Kim et al. (1996).

The deviation of the local Hubble expansion from the global one
would have influence on the calibration of the number counts
of distant sources. We have found that the number density of
distant faint galaxies would be underestimated  
if $H_L>H_G$. This may partially remove the demand for a strong
evolution scenario of galaxies population in order 
to account for the number excess of faint (blue) galaxies. 
As a result, the theoretically predicted number of giant arcs, which
are the distorted images of distant galaxies by the gravitational
potentials of foreground clusters of galaxies, would 
increases. This probably provides another explanation for
the deficit of giant arcs
in statistical lensing (Wu \& Mao 1996).

We have only discussed the effect of $H_L/H_G$ on the 
magnitude-redshift relation and the number counts. Indeed, 
if the local expansion is larger than the global one,
many aspects of astrophysics might need to be modified. 
The crucial point depends on the measurements of both
the Hubble constant and the matter density at different
redshift. The current status on the 
determinations of $H_0$ and $\Omega_0$ appears to be 
unfortunate: The observations using the same 
method can result in very different results.
Here we do not intend to be involved in the 
disputes. Instead, we point out that it would be of
great significance if one would explore other possibilities
which can be used to probe or 
constrain the variations of the Hubble constant
and the density parameter.

\acknowledgments

We are grateful to the referees, Allan Sandage and David Branch, 
for their constructive comments and suggestions.  
WXP was supported by the National Science Foundation
of China and a World-Laboratory fellowship.

\clearpage

\figcaption{Limit on the variation of the Hubble
	constant $H_L/H_G$ against the global mass 
        density $\Omega_0$ using the distant type Ia supernova 
	at $z=0.458$. The solid line corresponds to the 
	peak magnitude $m_R=22.2$ and the dashed
	lines are the error bars arising from the intrinsic 
	dispersion ($\pm0.25$) of the local type Ia supernovae 
	and the measurement 
	uncertainties at maximum light ($\pm0.16$)
	(The supernova data are taken from Perlmutter et al. 1995).
             \label{fig1}}

\figcaption{Differential number counts of galaxies in $B$
	band per square degree for an uniform Hubble
	flow $H_L=H_G$ and a variable Hubble constant 
	$H_L=1.3H_G$, respectively.
        Evolution effect has not been included.
             \label{fig2}}

\figcaption{Variations of the Hubble expansion rate 
      ($H_L/H_G$) against the global mean mass density of 
	the universe $\Omega_0$ for a given local 
	mass density $\Omega_{L}$. 
 \label{fig3}}

\epsfbox{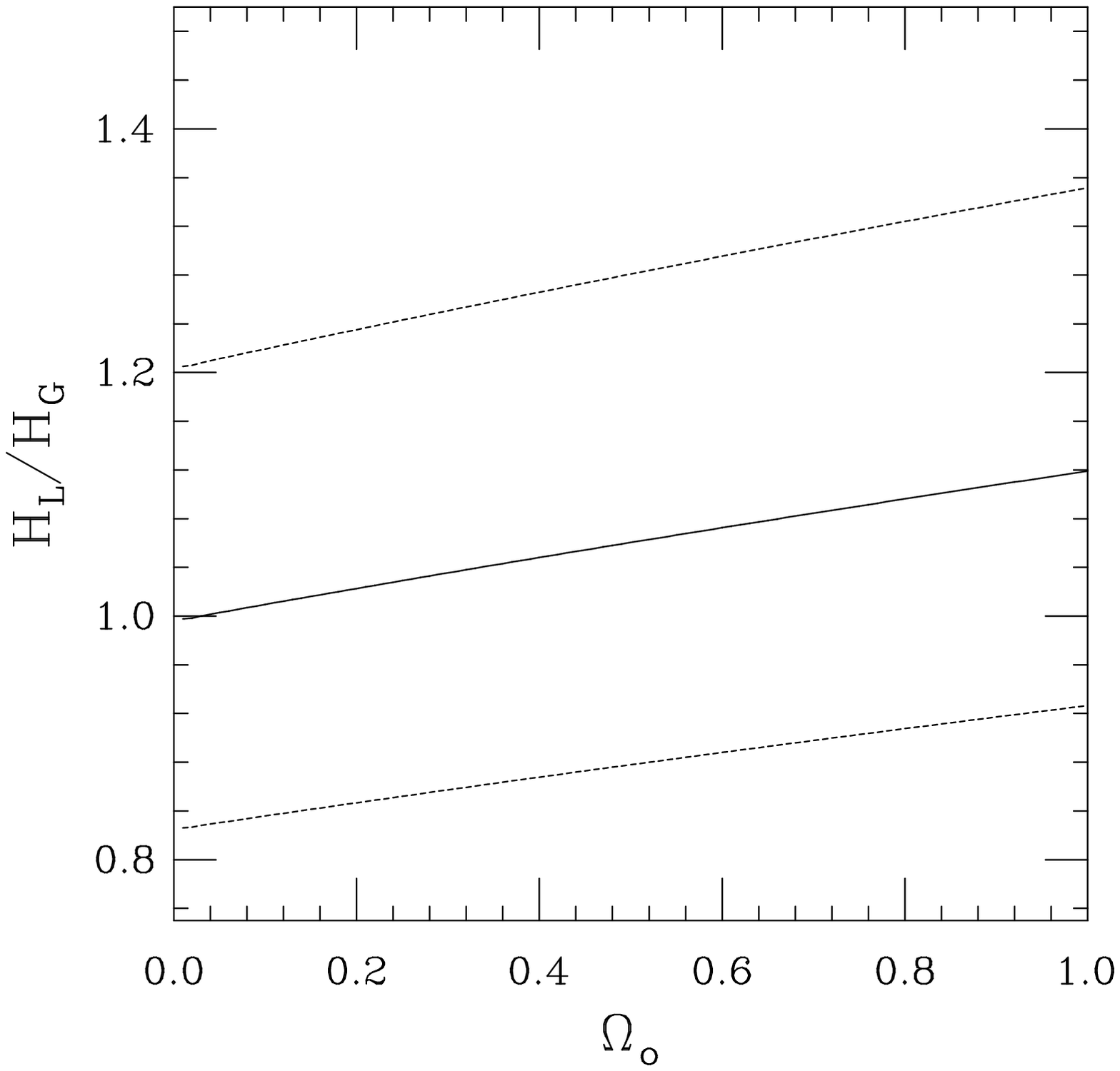}

\epsfbox{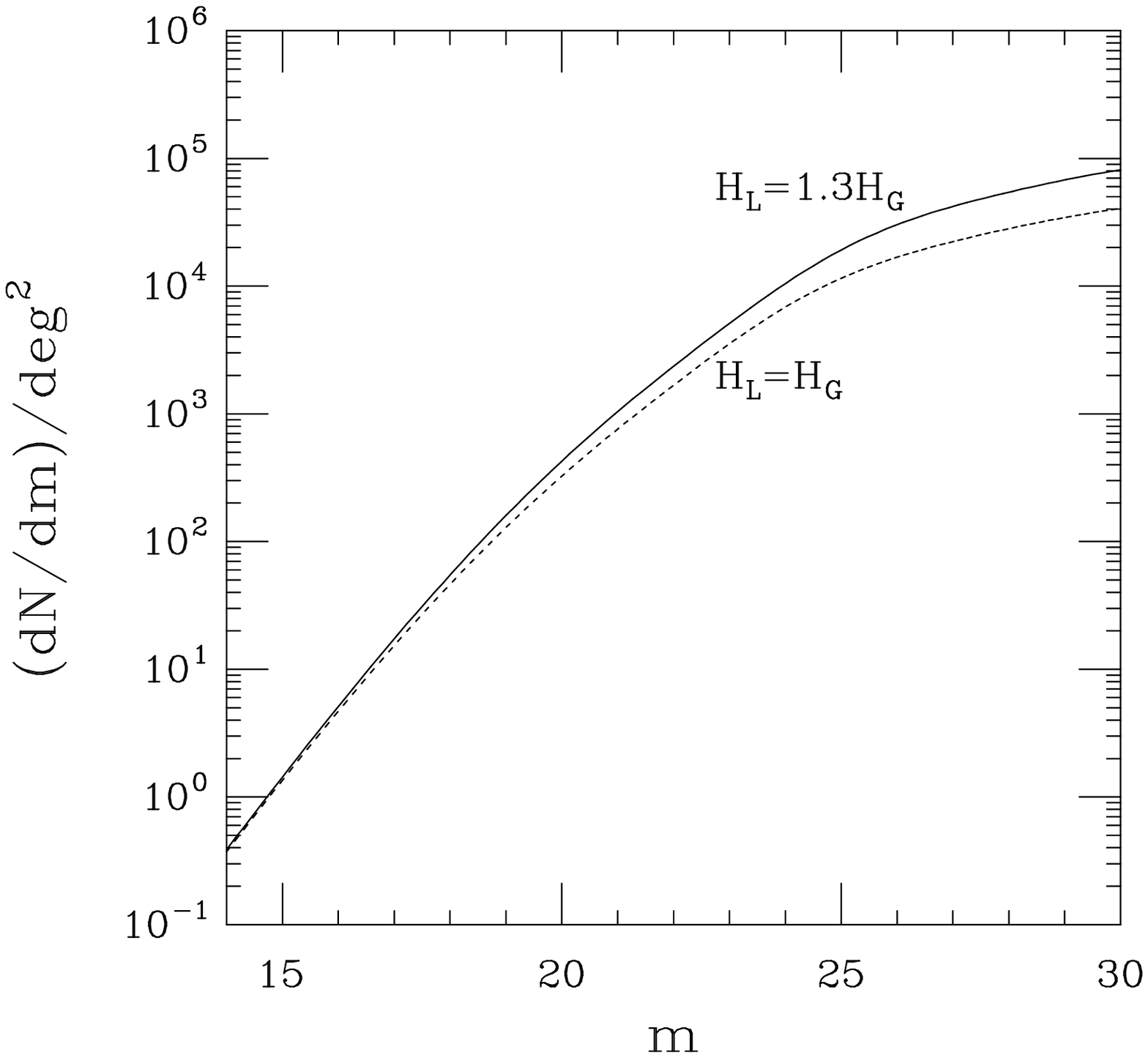}

\epsfbox{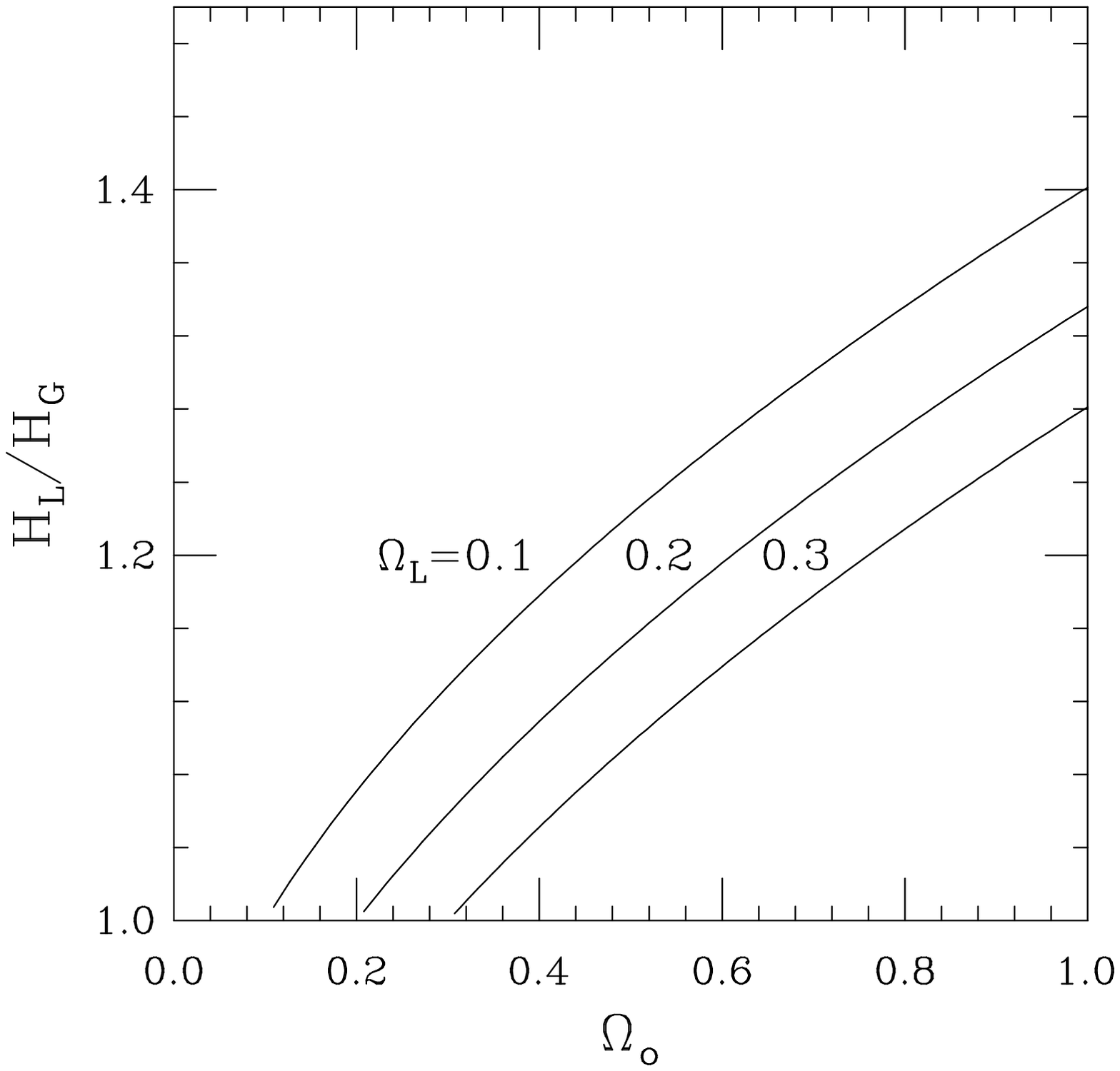}


\clearpage


\begin{references}
\reference{}Bahcall, N. A., Lubin, L. M., \& Dorman, V. 1995, \apj, 447, L81
\reference{}Bennett, C. L., et al. 1996, \apj, submitted
\reference{}Birkinshaw, M., \& Hughes, J. P. 1994, \apj, 420, 33
\reference{}Bolte, M., \& Hogan, C. J. 1995, \nat, 376, 399
\reference{}Branch, D., Nugent, P., \& Fisher, A. 1996,
	Proc. of NATO ASI on Thermonuclear Supernovae, eds. 
	R. Canal, P. Ruiz-Lapuente \& J. Isern (Dordrecht: Kluwer), 
	in press 
\reference{}Branch, D., \& Tammann, G. A. 1992, \araa, 30, 359
\reference{}Broadhurst, T. J., Ellis, R. S., \& Glazebrook, K. 1992,
		\nat, 355, 55
\reference{}Efstathiou, G., Ellis, R. S., \& Peterson, B. A. 
	    1988, \mnras, 232, 431
\reference{}Fort, B., \& Mellier, Y. 1994, \aapr, 5, 239
\reference{}Freedman, W. L., et al. 1994, \nat, 371, 757
\reference{}Goobar, A., \& Perlmutter, S. 1995, \apj, 450, 14
\reference{}Grogin, N. A., \& Narayan, R. 1996, \apj, submitted
\reference{}Herbig, T., Lawrence, C. R., \& Readhead, A. C. S.
		1995, \apj, 449, L5
\reference{}Inagaki, Y., Suginohara, T., \& Suto, Y. 1995,
	    \pasj, 47, 411
\reference{}Jones, M., et al. 1993, \nat, 365, 320
\reference{}Kim, A. et al. 1996, Proc. of NATO ASI on Thermonuclear
	Supernovae, eds. R. Canal, P. Ruiz-Lapuente \& J. Isern 
	(Dordrecht: Kluwer), in press 
\reference{}King, C. R., \& Ellis, R. S. 1985, \apj, 288, 456
\reference{}Lauer, T. R., \& Postman, M. 1992, \apj, 400, L47
\reference{}Nakamura, T. T., \& Suto, Y. 1995, \apj, 447, L65
\reference{}Ostriker, J. P., \& Steinhardt, P. J. 1995, \nat, 377, 600 
\reference{}Perlmutter, S., et al. 1995, \apj, 440, L41
\reference{}Perlmutter, S., et al. 1996, Proc. of NATO ASI on Thermonuclear
	Supernovae, eds. R. Canal, P. Ruiz-Lapuente \& J. Isern 
	(Dordrecht: Kluwer), in press 
\reference{}Pierce, M. J., et al. 1994, \nat, 371, 385 
\reference{}Rephaeli, Y. 1995, \araa, 33, 541
\reference{}Sandage, A. 1996, private communication
\reference{}Sandage, A., \& Hardy, E. 1973, 183, 743
\reference{}Sandage, A., \& Tammann, G. A. 1996, ApJ, submitted
\reference{}Sandage, A., Tammann, G. A., \& Hardy, E. 
	     1972, \apj, 172, 253
\reference{}Shi, X., Widrow, L. M., \& Dursi, L. J. 1995, preprint
\reference{}Sunyaev, R. A., \& Zel'dovich, Y. B. 1972, Comm. Astrophys.
  	Space Sci., 4, 173
\reference{}Suto, Y., Suginohara, T., \& Inagaki, Y. 1995,
	   Prog. Theor. Phys. Phys., 93, 839 
\reference{}Tammann, G. A., \& Sandage, A. 1995, ApJ, 452, 16
\reference{}Tanvir, N. R., Shanks, T., Ferguson, H. C., \& Robinson, D. R. T.
 	1995, \nat, 377, 27
\reference{}Turner, E. L., Cen, R., \& Ostriker, J. P. 1992, \aj, 103, 1427
\reference{}Turner, E. L., Ostriker, J. P., \& Gott III, H. R.
	1984, \apj, 284, 1
\reference{}Whitmore, B. C., Sparks, W. B., Lucas, R. A., Macchetto, F. D.,
	\& Biretta, J. A. 1995, \apj, 454, L73
\reference{}Wu, X. P. 1996, Fund. of Cos. Phys., in press
\reference{}Wu, X. P., Deng, Z., Zou, Z., Fang, L. Z., \& Qin, B. 
	1995, \apj, 448, L65
\reference{}Wu, X. P., \& Hammer, F. 1993, \mnras, 262, 187
\reference{}Wu, X. P., \& Mao, S. 1996, \apj, in press
\reference{}Yamashita, K. 1994, New Horizon of X-Ray Astronomy: First Results 
	from {\it ASCA}, eds. F. Makino \& T. Ohashi (Tokyo: 
	Universal Academy Press), 279
\reference{}Yoshii, Y., \& Sato, K. 1992, \apj, 387, L7
\end{references}
\end{document}